\documentclass[jmp,%
 aip,
 amsmath,amssymb,
 reprint,%
]{revtex4-1}

\usepackage{graphicx}
\usepackage{dcolumn}
\usepackage{bm}
\usepackage{hyperref}
\usepackage{moresize}
\usepackage{amsmath}
\usepackage[utf8, applemac]{inputenc}
\usepackage[T1]{fontenc}
\usepackage{mathptmx}
\usepackage{etoolbox}
\usepackage{amssymb}
\usepackage{xcolor}

\definecolor{1purple}{RGB}{85,63,150}
\definecolor{1blue}{RGB}{8,57,148}
\definecolor{1pink}{RGB}{249,211,189}
\definecolor{1wheat}{RGB}{196,196,27}
\definecolor{1grey}{RGB}{187,187,187}
\definecolor{1red}{RGB}{231,20,26}
\definecolor{1green}{RGB}{33,146,75}
\definecolor{1cyan}{RGB}{111,199,213}
\renewcommand{\eqref}[1]{\mbox{Eq.~(\ref{#1})}} 
\newcommand{\figpanel}[2]{Fig.~\hyperref[#1]{\ref*{#1}(#2)}} 
\newcommand{\figpanels}[3]{Fig.~\hyperref[#1]{\ref*{#1}(#2)-(#3)}} 
\newcommand{\figpanelNoPrefix}[2]{\hyperref[#1]{\ref*{#1}(#2)}} 
\newcommand{\figpanelsNoPrefix}[3]{\hyperref[#1]{\ref*{#1}(#2)-(#3)}} 

\usepackage{mleftright} 

\usepackage{units}
\hypersetup{%
	plainpages=true,
	breaklinks=true,
	hypertexnames=false,
	pageanchor=true,
	colorlinks=true,
	linkcolor={blue},
	citecolor={blue},
	urlcolor={blue},
	anchorcolor={black}
}
\usepackage[all]{hypcap} 

\makeatletter
\def\@email#1#2{%
 \endgroup
 \patchcmd{\titleblock@produce}
  {\frontmatter@RRAPformat}
  {\frontmatter@RRAPformat{\produce@RRAP{*#1\href{mailto:#2}{#2}}}\frontmatter@RRAPformat}
  {}{}
}%
\makeatother
\begin{document}

\preprint{AIP/123-QED}

\title{Slowing and Storing Microwaves in a Single Superconducting Fluxonium Artificial Atom} 



\author{Ching-Yeh Chen}
\affiliation{Department of Physics, National Tsing Hua University, Hsinchu 30013, Taiwan}

\author{Shih-Wei Lin}
\affiliation{Department of Physics, National Tsing Hua University, Hsinchu 30013, Taiwan}

\author{Ching-Ping Lee}
\affiliation{Department of Physics, National Tsing Hua University, Hsinchu 30013, Taiwan}

\author{J.~C.~Chen}
\affiliation{Department of Physics, National Tsing Hua University, Hsinchu 30013, Taiwan}

\author{I.-C.~Hoi}
\affiliation{Department of Physics, National Tsing Hua University, Hsinchu 30013, Taiwan}
\affiliation{Department of Physics, City University of Hong Kong, Kowloon, Hong Kong SAR 999077, China}

\author{Yen-Hsiang Lin}
\affiliation{Department of Physics, National Tsing Hua University, Hsinchu 30013, Taiwan}
\affiliation{Taiwan Semiconductor Research Institute, Hsinchu 300091, Taiwan}

\date{\today}


\begin{abstract}

Three-level $\Lambda$ systems provide a versatile platform for quantum optical phenomena such as Electromagnetically Induced Transparency (EIT), slow light, and quantum memory. Such $\Lambda$ systems have been realized in several quantum hardware platforms including atomic systems, superconducting artificial atoms, and meta-structures. Previous experiments involving superconducting artificial atoms incorporated coupling to additional degrees of freedom, such as resonators or other superconducting atoms. In this work, we performed an EIT experiment in microwave frequency range utilizing a single Fluxonium qubit within a microwave waveguide. The $\Lambda$ system is consisted of two plasmon transitions in combination with one metastable state originating from the fluxon transition. In this configuration, the controlling and probing transitions are strongly coupled to the transmission line, safeguarding the transition between $|0\rangle$ and $|1\rangle$ states, and ensuring the Fluxonium qubit is close to the sweet spot. Our observations include the manifestation of EIT, a slowdown of light with a delay time of 217 ns, and photon storage. These results highlight the potential as a phase shifter or quantum memory for quantum communication in superconducting circuits.
\end{abstract}

\pacs{}

\maketitle 




With the expansive progress in quantum information science, the interaction between light and matter\cite{frisk2019ultrastrong} has triggered very active research, with unique applications to the storage of the quantum information and quantum networks\cite{cirac1997quantum}. Electromagnetically induced transparency (EIT)\cite{RevModPhys.77.633,sheremet2023waveguide}, the controllable transparency caused by the interference between the three-level $\Lambda$ system, plays an important role in physics. Based on EIT, one can adeptly manipulate the deceleration of light and quantum memory\cite{lvovsky2009optical,bajcsy2009efficient}, thereby facilitating the interconnection of quantum nodes to construct a robust quantum internetwork\cite{kimble2008quantum,bhaskar2020experimental}. Furthermore, the nonlinearity of the EIT has been extensively investigated across various systems and frequency range, including natural atoms\cite{PhysRevLett.66.2593, PhysRevLett.84.5094, PhysRevLett.120.183602}, coupling cavities\cite{bao2021demand, liu2017electromagnetically}, meta-structures\cite{wang2019enhanced,lao2019dynamically}, ion trap\cite{feng2020efficient,jordan2019near} and superconducting circuits\cite{PhysRevLett.120.083602, PhysRevA.93.053838,brehm2022slowing}.

Recently, the exploration of microwave quantum optics using superconducting qubits coupling to the transmission line, also known as waveguide quantum electrodynamics(QED) is blossoming. Experimental demonstrations have showcased the advantages of robust qubit coupling to electromagnetic waves \cite{wen2018reflective,wilson2011observation, RevModPhys.89.021001}, with the characteristic of the qubits coupling to the continuous photon mode, making it possible to directly measure\cite{Nat.Commun.12.6383.(2021),kou2018simultaneous}, observe\cite{brehm2021waveguide,wen2019large}, and control\cite{cheng2024tuning,hoi2011demonstration}, inspiring the building block of quantum network\cite{kannan2023demand,gheeraert2020programmable,redchenko2023tunable} and realizing the all-to-all connection\cite{zhou2023realizing}. Nonetheless, constructing the three-level $\Lambda$ system is still difficult without considering other degrees of freedom (e.g. coupling cavity \cite{novikov2016raman, PhysRevResearch.5.033192}, qubits coupling\cite{chiang2022tunable} and interference of the giant atom\cite{PhysRevA.103.023710}). In waveguide QED, the strong light matter interaction and the intrinsic multi-level structure of superconducting qubits naturally provide the platform for constructing three-level $\Lambda$ systems, enabling phenomena such as EIT.\cite{abdumalikov2010electromagnetically,vadiraj2021engineering} However, using a single atom to construct the three-level $\Lambda$ system and demonstrate the EIT is still an unsolved but important problem, which is not only simplifying the degree of freedom and reducing fabrication complexity, but also provides a clean and scalable platform to explore quantum interference at the few-photon level. With the benefit of the strong coupling of artificial atom, the truly single atom interacting with light can come true and easier to understand the interaction of the EIT.

In this paper, we present the first observation of slowing light and quantum memory based on EIT by constructing the $\Lambda$ system with a single artificial atom without considering any degree of freedom. Our feature is based on the theory of the fluxonium qubit\cite{fluxoniumQ}  in the waveguide QED, and we create the $\Lambda$ system with the advantage of the Plasmon transition and the metastable state from the fluxon transition of the lowest transition as we control the external flux $\phi_{\rm{ext}}$ close to the sweet spot.   Our approach involves a two tone process for driving on the transition frequency of the $\Lambda$ system.
By introducing the coupling field, as in real atomic systems, the EIT phenomenon is observed and the magnitude response of the probe light is significantly changed and becomes much sharper.
In time domain, we observed the slowing down light with delay time $\tau_{\rm{d}}= 217 \,\, \rm{ns}$ by sending a gaussian pulse. Moreover, we successfully stored and retrieved the microwave pulse, characterized by an average photon count below the single photon level, in the lowest state of the fluxonium qubit by dynamically modifying the shape of the coupling field with the storage efficiency close to $12\,\,\%$.

\begin{figure*}
    \includegraphics[scale=0.44]{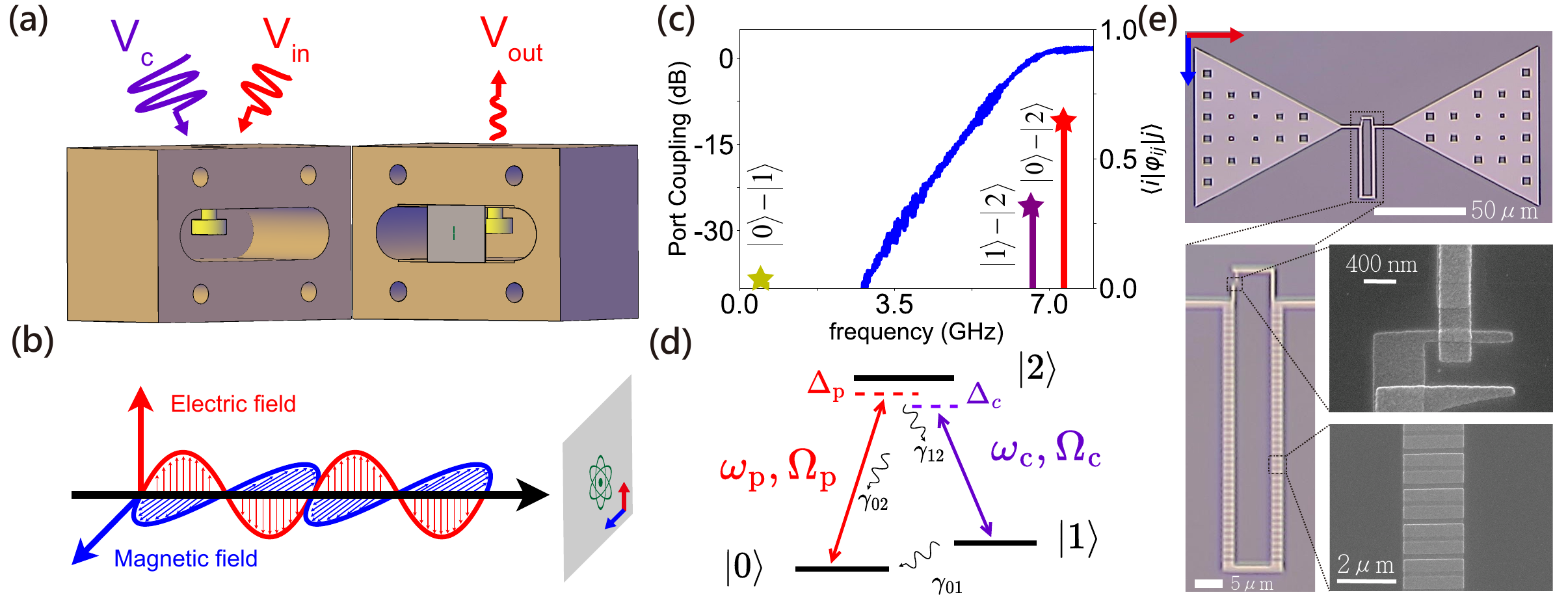}
    \caption{Light matter interaction of the Fluxonium qubit implemented in a effective one-dimensional waveguide. (a) Experimental arrangement showing the fluxonium qubit inside the copper three-dimensional (3D) waveguide. The 3D waveguide has the cut-off frequency at $6.5 \,\,\rm{GHz}$, shown in (c). We send in the input field, $V_{\rm{in}}$, and control field, $V_{\rm{c}}$, through one of the port, and measure the $V_{\rm{out}}$. (b) Schematic diagram of a single Fluxonium qubit interacting with the electromagnetic field. The microwave field is confined as $TE_{10}$ mode of the waveguide and the electric field is polarized in one direction.  (c) Transition frequencies and corresponding $\langle i|\varphi_{ij}|j\rangle$ of the Cooper pair number operator $-i\partial_{\phi}$ (stars), and we bias at $\phi_{ext}/\phi_0$=0.53. The solid blue curve shows microwave transmission measured with a two-port configuration at room temperature.  (d) The schematic diagram of the lowest three energy levels of the fluxonium qubit. The three-level system is driven by a control field (purple) and a weak probe field (red), where the $\Omega_{p} \ll \Gamma_{02}/2 + \gamma_{22}$. The probe field with frequency $\omega_{\rm{p}}$ has a detuning $\Delta_{\rm{p}}$ with the transition between $|0\rangle$and $|2\rangle$ and the control field with frequency $\omega_{\rm{c}}$ has a detuning $\Delta_{\rm{c}}$ with the transition between $|1\rangle$ and $|2\rangle$. The corresponding driving strengths are $\Omega_{\rm{p}}$ and $\Omega_{\rm{c}}$.(e) Optical microscope images of the measured device. The bowtie antenna connects to the weak junction and junction arrays of the fluxonium. The holes of the antenna are dug to avoid the formation of mobile vortex causing the frequency fluctuation of the qubit. The red and blue arrows indicate the polarization direction of the electric and magnetic field respectively. The SEM images show the details of the weak junction and the part of the junction array as a superinductance.}
    \label{sample}
\end{figure*}

Our setup is a single fluxonium circuits implemented in a three-dimensional(3D) copper waveguide, as shown in Fig. \ref{sample}(a). The microwave in the passband of the waveguide propagates in one dimension, and penetrates through the thin film of the fluxonium circuits. The electric field of the propagating microwave is polarized within the waveguide's $TE_{10}$ mode. Consequently, the dipole-dipole interaction direction is perpendicular to the light propagation direction, as illustrated in Fig. \ref{sample}(b).
 Our artificial atom, a fluxonium circuit, consists of a solitary Josephson junction with Josephson tunneling energy $E_{J}$ connected in parallel to a capacitance C and inductance L, with charging energy $E_{c} = e^{2}/2C$ and the inductive energy $E_{L} = (\hbar/2e)^{2}/L$ respectively. To operate in the typical artificial fluxonium atom, these parameters must meet two key conditions: $E_{L} \ll E_{J}$ and $E_{C} \ll E_{J}$\cite{fluxoniumQ,kou2018simultaneous}.  To enhance the difference of the decay rate between the plasmon transition and the fluxon transition, we exposed the plasmon transitions ($|0\rangle \leftrightarrow |2\rangle$ and $|1\rangle \leftrightarrow |2\rangle$) in the passband in our waveguide, and placed the fluxon transition ($|0\rangle \leftrightarrow |1\rangle$) in the stop band with 40 dB isolation, as shown in Fig. \ref{sample}(c), comparing with passband.

\begin{figure}
    \includegraphics[scale=0.24]{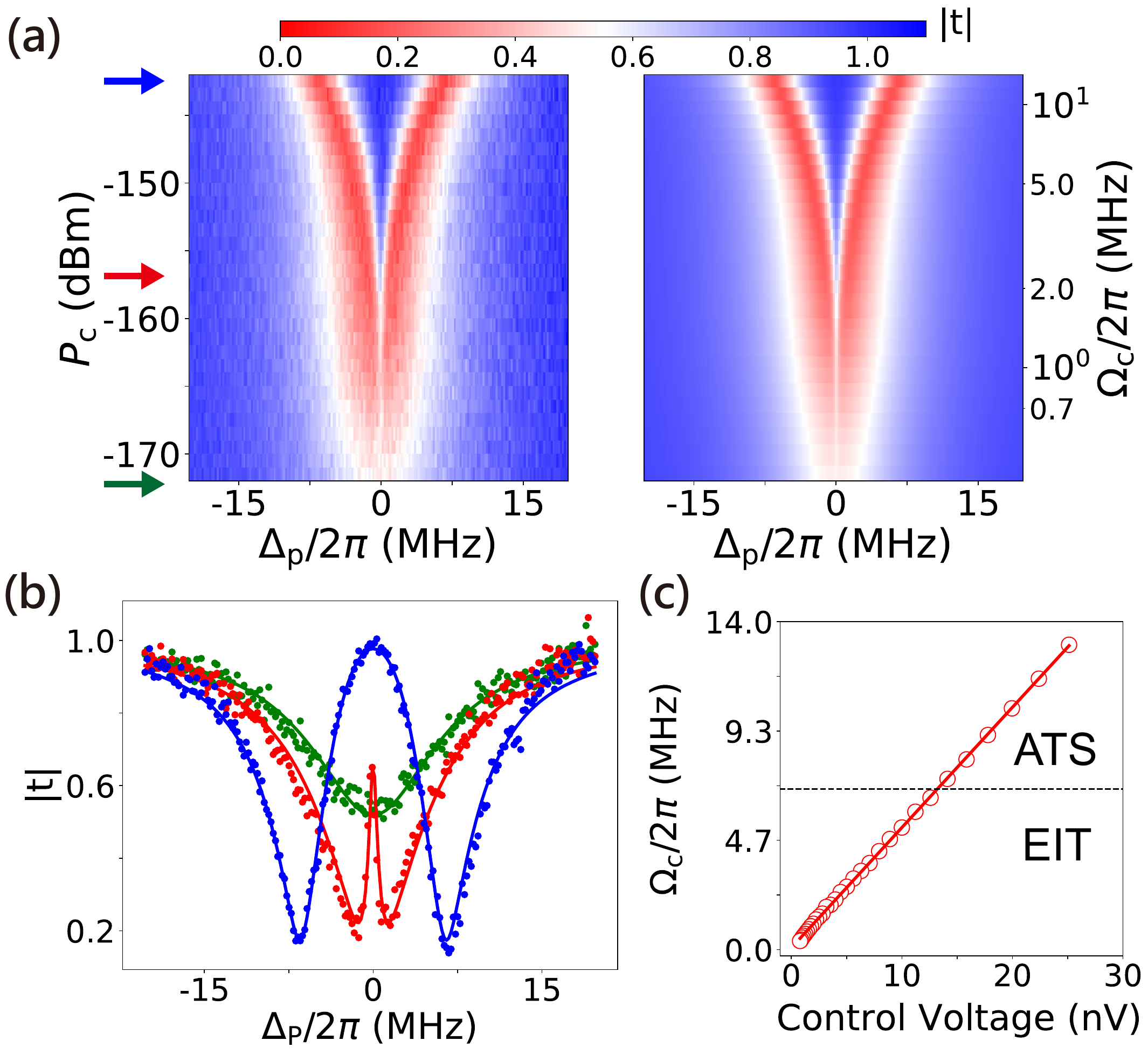}
    \caption{Spectroscopy of EIT. (a) Transmission coefficient $|t|$ as a function of power of control tone field $P_{\mathrm{c}}$ and $\Delta_{\mathrm{p}}$ for both experimental data and fitting results obtained using Eq.~\ref{mseq}. (b) Linecuts of the experimental data and corresponding fits from (a), the correspoding power is indicated by the arrows in (a), illustrating the crossover from EIT to ATS for different control field strengths: $P_{\mathrm{c}}=-142\,\mathrm{dBm}$ (blue),$-157\,\,\mathrm{dBm}$ (red), and$-172\,\,\mathrm{dBm}$ (green). (c) Effective $\Omega_{\mathrm{c}}$ as a function of control voltage, extracted by fitting the measured transmission coefficient $|t|$. The linear fit demonstrates $\Omega_{\mathrm{c}}\propto\sqrt{P_{\mathrm{c}}}$. The black dashed line represents the threshold for $\Omega_{\mathrm{EIT}}$.
}
    \label{EIT}
\end{figure}

Figure. \ref{sample}(d) shows the $\Lambda$ system energy levels. The system is driven with the $\omega_{\rm{c}}$ near the $|1\rangle \leftrightarrow |2\rangle$ transition $\omega_{\rm{12}}$ with a microwave control tone of amplitude (Rabi-strength) $\Omega_{\rm{c}}$. The $\omega_{\rm{p}}$ near the $|0\rangle \leftrightarrow |2\rangle$ transition $\omega_{\rm{02}}$ with a microwave probe tone of amplitude (Rabi-strength) $\Omega_{\rm{p}}$. Here we introduce the detuning $\Delta_{\rm{p}}=\omega_{\rm{p}} - \omega_{02}$ and $\Delta_{\rm{c}}=\omega_{\rm{c}} - \omega_{12}$. We use the two-tone spectroscopy of the shifting from these levels hybridize and split into two dressed states separated by $\Omega_{\rm{c}}$, forming the ATS\cite{brehm2022slowing} to calibrate the power of the control field $P_{\rm{c}}$ in Fig. \ref{EIT}(a). 

Our device, shown in Fig. \ref{sample}(e), is manufactured on a silicon (Si) substrate using the Dolan bridge and angle deposition process\cite{DolanBridge}. It features an Al/AlOx/Al small junction with dimensions measuring 150 nm $\times$ 270 nm. Additionally, a superinductance\cite{superinductance} is implemented through an array of 180 large-area Josephson junctions. These two components together form the loop, enabling the tuning of the transition energy by manipulating $\phi_{\rm{ext}}$. Once these conditions are met, we derive the Hamiltonian\cite{PhysRevLett.103.217004} for the fluxonium circuits.
\begin{equation}
{H_{f}} = 4{E_{C}}{n^{2}} + \frac{1}{2}{E_{L}}{\phi^{2}} - {E_{J}}{\cos(\phi + {\phi_{\rm{ext}})}}
\label{Hami}
\end{equation}
Where $\phi$ represents the phase twist across the inductance and $2e\times n$ signifies the displacement charge at the capacitance, these two operators obey the commutation relation $[\phi,n]=i$. The parameter ${\phi_{\rm{ext}}}$ represents the reduced magnetic flux biasing the loop. At ${\phi_{\rm{ext}}/\phi_{0}}=0.5$, this biasing point is referred to as the "sweet spot," where the system exhibits high coherence for the lowest transition. 

Here, we measured the transmission coefficient, $t=\langle V_{out} \rangle/\langle V_{in} \rangle$ with applying the input field \(V_{\rm{in}}\) through a heavily attenuated and filtered microwave line to the 3D waveguide, and detecting the output signal \(V_{\rm{out}}\). To characterize the fluxonium circuits parameters,  we measured the one-tone spectroscopy as a function of $\phi_{\rm{ext}}$, and the theoretical fitting curves are presented in Fig. S2\cite{SuppMaterial}. The fitting line represents the outcome of numerically diagonalizing Hamiltonian (1) with $E_{J}=9.041$ GHz, $E_{C}=0.995$ GHz, $E_{L}=0.807$ GHz. For creating the $\Lambda$ system, we operate the artificial fluxonium atom near the sweet spot ($\phi_{\rm{ext}}/\phi_0 = 0.53$). The states $|0\rangle$ and $|1\rangle$ are still like the tunnel splitting of the two-fold degenerate classical ground state, but the transitions to the state $|2\rangle$ are dipole-allowed.

The dynamics of the system can be described by the following Born-Markov master equation.

\begin{equation}
\begin{split}
{\dot{\rho}} =&-i[\widetilde{H},\rho] + \Gamma_{02}\mathcal{D}[\sigma_{02}]\rho+ \Gamma_{12}\mathcal{D}[\sigma_{12}]\rho + \Gamma_{01}\mathcal{D}[\sigma_{01}]\rho +\\ &\Gamma_{10}\mathcal{D}[\sigma_{10}]\rho + 2\gamma_{22}\mathcal{D}[\sigma_{22}]\rho+ 2\gamma_{11}\mathcal{D}[\sigma_{11}]\rho
\label{mseq}
\end{split}
\end{equation}

Here, $\mathcal{D}$ denotes a superoperator\cite{Lindblad1976} defined as $\mathcal{D}[A]\rho=2A\rho A^{\dag}-A A^{\dag}\rho +\rho A A^{\dag}$, where $A$ represents the operator $\sigma_{ij}$, with $i \leq j$, and ${i,j}\in {0,1,2}$. The coefficient within $\mathcal{D}[A]\rho$ signifies the dissipation rate, encompassing both relaxation rates $\Gamma_{ij}$, pure dephasing rates $\gamma_{jj}$, and we can have the decoherence rate $\gamma_{ij}=\Gamma_{ij}/2+\gamma_{jj}$. Given that $\omega_{01}/2\pi$ is 648 $\rm{MHz}$, considering the temperature affecting according to the Boltzmann distribution\cite{PhysRevLett.130.267001}. We introduce $\Gamma_{01}$ to account for thermal excitation, as it significantly impacts this transition. However, the higher frequencies, $\omega_{21}$ and $\omega_{20}$, stand sufficiently above the thermal scale, enabling us to disregard their thermal effects\cite{wiegand2021ultimate}. $\widetilde{H}$ is the system Hamiltonian, described in an appropriate rotating frame.

\begin{equation}
\begin{split}
\widetilde{H} = &-\frac{1}{2}\Omega_{p}(\sigma_{02}+\sigma_{20})-\frac{1}{2}\Omega_{c}(\sigma_{12}+\sigma_{21})+\\&(\Delta_{p}-\Delta_{c})\sigma_{11}+\Delta_{p}\sigma_{22}
\end{split}
\end{equation}

\begin{table*}
\begin{tabular}{|c|c|c|c|c|c|c|c|c|c|}

\hline
$\omega_{\rm 02}/2\pi$ & $\omega_{\rm 12}/2\pi$ & $\omega_{\rm 01}/2\pi$ & $\Gamma_{\rm 02}/2\pi$ & $\Gamma_{\rm 12}/2\pi$ & $\Gamma_{\rm 01}/2\pi$ & $\Gamma_{\rm 10}/2\pi$& $\gamma_{\rm 11}/2\pi$ & $\gamma_{\rm 22}/2\pi$&$\varphi$\\
\hline
$\rm{GHz}$ & $\rm{GHz}$ & $\rm{MHz}$ & $\rm{MHz}$ & $\rm{MHz}$ & $\rm{MHz}$ & $\rm{MHz}$ & $\rm{MHz}$& $\rm{MHz}$& $rad$\\
\hline
$7.329$ & $6.681$ & $648$ & $13.78$ & $2.08$ & $0.022$& $0.0218$& $0.14$ & $0.16$ & -0.299\\
\hline
\end{tabular}
\caption{Extracted parameters of the three lowest transitions for the fluxonium when we bias at $\phi_{ext}/\phi_0=0.53$. We extract the anharmonicity between the $|0\rangle \leftrightarrow |2\rangle$, $|1\rangle \leftrightarrow |2\rangle$ from single-tone spectroscopy\cite{SuppMaterial} and two-tone spectroscopy (data not show), and we get $\omega_{01}$ from $\omega_{01} = \omega_{02}-\omega_{12}$. We extract the decay rate from Fig.\ref{EIT} (a). $\Gamma_{10}$ is considered to compensate for the influence of the temperature. $\varphi$ is the impedance mismatch \cite{lu2021propagating}in the waveguide.}
\label{tab:2}
\end{table*}

To understand the three-level system, we solve the master equation Eq.\ref{mseq} using the QuTip\cite{johansson2012qutip} and obtained the transmission coefficient by the input-output relation\cite{PhysRevA.88.043806} for the steady-state and the time dynamics results.
\begin{equation}
{{t} = {1+i\frac{\Gamma_{02}}{\Omega_{p}}\rho_{02}}}
\label{tc}
\end{equation}

In Fig. \ref{EIT}(a), we show the steady-state result of the EIT. We bias at $\phi_{ext}/\phi_{0}=0.53$ to construct the $\Lambda$ system, the parameters are shown in Table \ref{tab:2}. The transmission coefficient $t$ as function of the $\Delta_{\rm{p}}$ and the power of the control field $P_{\rm{c}}$ which we drive on the $\omega_{\rm{12}}$. We get the perfect fit between the experimental data and the theoretical result. The dip in the transmission signal becomes deeper with the evolution of $P_{\rm{c}}$, and the probe signal exhibits transparency at $\Delta_{\rm{p}} \approx 0$ MHz in the absence of power broadening, which shows the phenomenon of EIT. 
Figure. \ref{EIT}(b) shows the line-cut with 3 different $P_{\rm{c}}$, the change from weak control field(green) to the EIT regime(red), and ATS regime(blue). To clearly understand the threshold between the EIT and ATS regime\cite{PhysRevLett.124.240402}, we have eq.\ref{boundary}.
\begin{equation}
{\Omega_{EIT}}={\gamma_{02}-\gamma_{01}}
\label{boundary}
\end{equation}
For $\Omega_{\rm{c}} < \Omega_{\rm{EIT}}$ the system is in the EIT regime, and for $\Omega_{\rm{c}} > \Omega_{\rm{EIT}}$ the system is in the ATS regime. The result is shown in the Fig. \ref{EIT} (c). The difference between the EIT and the ATS, as determined using the Akaike information criterion (AIC)\cite{PhysRevA.103.023710} for their models, is further discussed in the Supplementary material.\cite{SuppMaterial}.\

\begin{figure*}
    \includegraphics[scale=0.38]{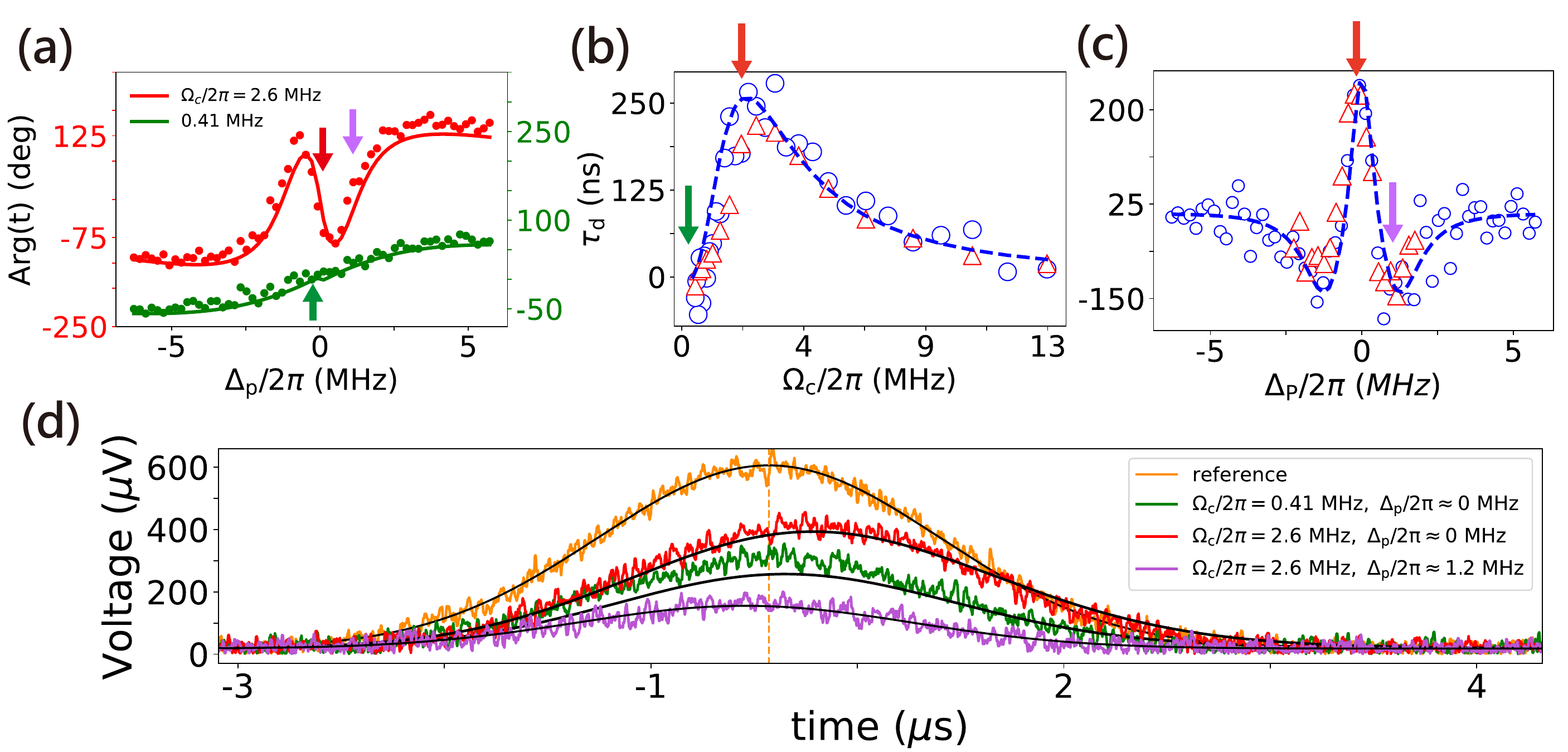}
    \caption{ Demonstration of slow and fast light. (a) Linecuts of $\mathrm{Arg(t)}$ as a function of $\Delta_{\mathrm{p}}$ from Fig.~\ref{EIT} (b) and (c) Delay time $\tau_{\mathrm{d}}$ as a function of $\Omega_{\mathrm{c}}$ and $\Delta_{\mathrm{p}}$, respectively. The experimentally extracted delay time $\tau_{\mathrm{d}}$ from pulsed measurements (red triangles), using the reference pulse for comparison, and spectroscopic measurements based on Eq.~\ref{delay}, are shown for d $\Delta_{\mathrm{p}}/2\pi\approx 0~\mathrm{MHz}$ and (d) $\Omega_{\mathrm{c}}/2\pi=2.6~\mathrm{MHz}$. Experimental data (blue circles) and theoretical results (blue dashed curve) are compared. The red and green arrows indicate slow light and absorption cases, respectively, while the purple arrow indicates the fast light case in (d). (d) Envelopes of the probe Gaussian pulse measured under two different $\Omega_{\mathrm{c}}$ and $\Delta_{\mathrm{p}}$. The slow light (red) and fast light (purple) cases are shown, in comparison with the reference pulse (orange), measured under far-detuned conditions, and the weak $\omega_{\rm{c}}$ case (green). The black solid lines are fits obtained by solving the master equation [Eq.~\ref{mseq}]. (b) (a), corresponding to the two different $\Omega_{\mathrm{c}}$ values indicated by the arrows of the same color in Fig.~\ref{EIT}(a). The solid curves show theoretical simulations. }

    \label{slow_light}
\end{figure*}

By introducing the control field, we can control the change of the phase with the EIT phenomenon, which indicates the reduction of the group velocity, shown in Fig. \ref{slow_light}(a). The green data is the typical phase response of the zoom-in data in Fig. \ref{EIT}(b). When $\Omega_{\rm{c}}=2.6 \,\, \rm{MHz}$, depicted as red data points, the slope of the phase is enlarged and opposite around $\Delta_{\rm{p}}/2\pi=0$, comparing with the typical $Arg(\rm{t})$. 
\begin{equation}
{\tau_{\rm{d}}}={-\frac{\partial Arg(t)}{\partial \omega_p}}
\label{delay}
\end{equation}
The Eq.\ref{delay} can be used to extracted the behavior of the $\tau_{\rm{d}}$ in the steady-state, as shown in the Fig. \ref{slow_light}(b) and (c). The maximum value of the $\tau_{\rm{d}}$ is indicated by the red arrows.

To demonstrate the $\tau_{\rm{d}}$ in time dynamics, we send a pulse as a Gaussian function $G(t,\sigma)$, where $G(t,\sigma)=\Omega_{\rm{p}}\exp{(-t^2/2\sigma^2)}$. We choose $\sigma=1\mu s$ and the weak probe amplitude $\Omega_{\rm{p}}$. When the $\Delta_{\rm{p}}/2\pi\approx0 \,\,\rm{MHz}$, the carrier frequency of the Gaussian pulse is $\omega_{\rm{p}}/2\pi=7.32625 \,\,\rm{MHz}$, which is applied to achieve impedance matching\cite{lu2021propagating}. We sweep the $\Omega_{\rm{c}}$ with a continuous control field, as shown in Fig. \ref{slow_light}(d). The green data in Fig. \ref{slow_light}(d) is the weak control field result, which is the interference result of the input field and the artificial fluxonium atom emission, and is indicated by the green arrow in Fig. \ref{slow_light}(a). We observed the largest $\tau_{\rm{d}}=217 \,\,\rm{ns}$ with $\Omega_{\rm{c}}/2\pi=2.6\,\, \rm{MHz}$ comparing with the reference pulse (orange)\cite{brehm2022slowing}. We also observe the fast light\cite{cheng2025group} as the $\Omega_{\rm{c}}/2\pi=2.6\,\, \rm{MHz}$ and $\Delta_{\rm{p}}/2\pi\approx 1.2 \rm{MHz}$, where the slope of the phase becomes opposite, indicated by the purple arrow in Fig. \ref{slow_light}(a) and the Fig. \ref{slow_light}(c).

\begin{figure}
    \includegraphics[scale=0.5]{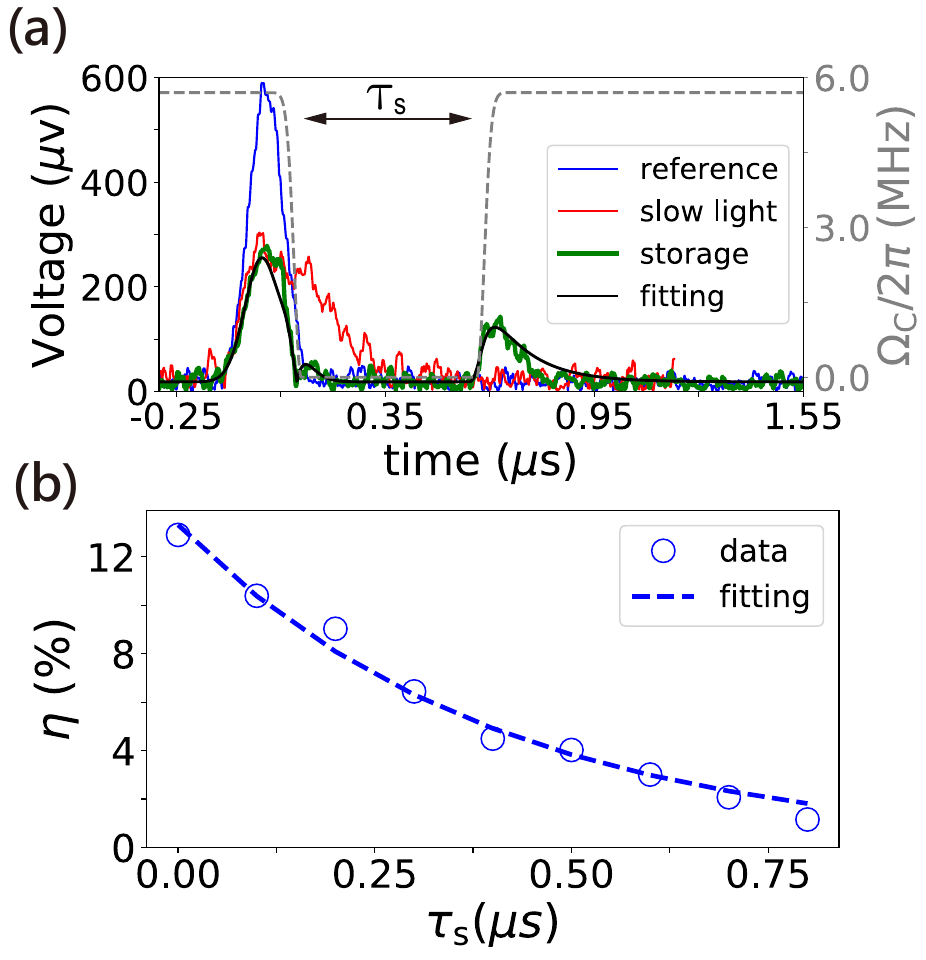}
    \caption{EIT-based microwave photon storage and retrieval (a)Storge pulse with the single atom. The reference pulse (blue) and the slow light (red).  The dynamical control pulse is shown as grey dashed curve. The green trace shows the stored-and-retrieval after $\tau_{\rm{s}}=0.5\,\, \mu s$. The black solid curve is the fitting result. (b) Storage efficiency $\eta$ as function of $\tau_{\rm{s}}$. The blue dashed line is the fitting of the experimental data.}
    \label{ps}
\end{figure}

 Next we demonstrate the photon storage experiment. We chose a shorter Gaussian probe pulse compared to the previous experiment. As shown in Fig. \ref{ps}(a), we input the Gaussian pulse (blue) with $\sigma_s=0.05 \mu s$, and the distortion of the slow light (red) is due to the narrow bandwidth of the EIT window\cite{chu2025slow}. The input average photon number $\langle N\rangle=\int G(t,\sigma)/(\sqrt{2\Gamma_{02}}) dt\approx 0.006$. Although the delay time value is smaller than the previous experiment, we still obviously distinguish the reference pulse and the slow light result. In this experiment, we dynamically turn off the control field between $\Omega_{\rm{c}}/2\pi=5.7 \,\, \rm{MHz}$ and $0 \,\,\rm{MHz}$. $\tau_s$ is the time When we start to store the photon in the artificial fluxonium atom and then turn on the control field to retrieve the photon back to the transmission line. We successfully stored the photon into the artificial fluxonium atom and retrieved it with $\tau_{\rm{s}}=0.5 \,\,\rm{\mu s}$. Our storage energy efficiency $\eta$ is defined as the ratio of the retrieved pulse and the reference pulse. In Fig. \ref{ps}(b), the maximum $\eta$ is up to $12\%$ and the dashed curve is the fitting with exponential decay.

The slowing down of group velocity observed in our results is mainly limited by $\gamma_{01}$; in the ideal case, $\tau_{\rm{d}}$ could be improved to 604 ns, analyzed from the spectroscopy data and eq.\ref{delay}. This limitation arises because we operate our artificial fluxonium atom at the non-sweet spot to construct the $\Lambda$ system. This issue could be solved by controlling the fluxonium atom's parameters to operate in a different regime\cite{mencia2024integer,ardati2024using}, enabling the construction of the $\Lambda$ system directly at the sweet spot.
 

In conclusion, we construct the $\Lambda$ system using a single superconducting fluxonium artificial atom and verify it through the EIT phenomenon. Comparing it with the reference\cite{chu2025slow}, we build up our system without any other degree of freedom between the qubit and the resonator. Our analysis reveals a close agreement between experimental data and fitting results, allowing us to extract a threshold value for $\Omega_{\rm{EIT}}/2\pi=6.85 \,\,,\rm{MHz}$ and slowing down the light with $\tau_{\rm{g}}=217\,\,,\rm{ns}$, with the wave propagation direction perpendicular to the interaction direction. We on-demand control the shape of the control field, store the photon into the artificial fluxonium atom, and retrieve it back to the transmission line with the maximum $\eta=12 \%$. 
This work explores the possibility of investigating quantum optics phenomena in a three-level system within waveguide QED. Furthermore, it advances quantum information processing in superconducting circuits by utilizing the result as a component for a phase shifter or quantum memory.



\begin{acknowledgments}
\indent
We thank Prof. Ite Yu for the valuable discussions.  YHL is supported by the National Science and Technology Council in Taiwan with the grant NSTC 113-2119-M-007-008, NSTC 114-2119-M-007-011,  and by the Ministration of Education Yushan Young Scholar Fellowship. This work is supported by the National Center for Theoretical Sciences, the Higher Education Sprout Project, Top Research Centers in Taiwan Key Fields Program and Center for Quantum Science and Technology funded by the Ministry of Education in Taiwan.

\end{acknowledgments}


\section*{Author Declarations}
\vspace{-5mm}
\subsection*{Conflict of Interest} 
\vspace{-5mm}
The authors have no conflicts to disclose.

\subsection*{Author Contributions}
\textbf{C.-Y.~Chen}: Conceptualization (lead); Data curation (lead); Writing - original draft (equal); Writing - review \& editing (lead); Methodology (lead); Formal analysis (lead).
\textbf{S.~W.~Lin}: Conceptualization (equal). 
\textbf{C.~P.~Lee}: Conceptualization (equal). 
\textbf{J.-C. Chen}: Conceptualization (equal).
\textbf{I.-C. Hoi}: Conceptualization (equal).
\textbf{Y.-H.~Lin}: Conceptualization (equal); Supervision (lead); Writing - review \& editing (equal); Project administration (lead); Funding acquisition (equal).


\section*{Data Availability}
The data that support the findings of this study are available from the corresponding author upon reasonable request.

\section*{REFERENCES}

\bibliography{main}

\begin{thebibliography}{10}

\bibitem{frisk2019ultrastrong}
Anton Frisk~Kockum, Adam Miranowicz, Simone De~Liberato, Salvatore Savasta, and Franco Nori.
\newblock Ultrastrong coupling between light and matter.
\newblock {\em Nature Reviews Physics}, 1(1):19--40, 2019.

\bibitem{cirac1997quantum}
Juan~Ignacio Cirac, Peter Zoller, H~Jeff Kimble, and Hideo Mabuchi.
\newblock Quantum state transfer and entanglement distribution among distant nodes in a quantum network.
\newblock {\em Physical Review Letters}, 78(16):3221, 1997.

\bibitem{RevModPhys.77.633}
Michael Fleischhauer, Atac Imamoglu, and Jonathan~P. Marangos.
\newblock Electromagnetically induced transparency: Optics in coherent media.
\newblock {\em Rev. Mod. Phys.}, 77:633--673, Jul 2005.

\bibitem{sheremet2023waveguide}
Alexandra~S Sheremet, Mihail~I Petrov, Ivan~V Iorsh, Alexander~V Poshakinskiy, and Alexander~N Poddubny.
\newblock Waveguide quantum electrodynamics: collective radiance and photon-photon correlations.
\newblock {\em Reviews of Modern Physics}, 95(1):015002, 2023.

\bibitem{lvovsky2009optical}
Alexander~I Lvovsky, Barry~C Sanders, and Wolfgang Tittel.
\newblock Optical quantum memory.
\newblock {\em Nature photonics}, 3(12):706--714, 2009.

\bibitem{bajcsy2009efficient}
Michal Bajcsy, Sebastian Hofferberth, Vlatko Balic, Thibault Peyronel, Mohammad Hafezi, Alexander~S Zibrov, Vladan Vuletic, and Mikhail~D Lukin.
\newblock Efficient all-optical switching using slow light within a hollow fiber.
\newblock {\em Physical review letters}, 102(20):203902, 2009.

\bibitem{kimble2008quantum}
H~Jeff Kimble.
\newblock The quantum internet.
\newblock {\em Nature}, 453(7198):1023--1030, 2008.

\bibitem{bhaskar2020experimental}
Mihir~K Bhaskar, Ralf Riedinger, Bartholomeus Machielse, David~S Levonian, Christian~T Nguyen, Erik~N Knall, Hongkun Park, Dirk Englund, Marko Lon{\v{c}}ar, Denis~D Sukachev, et~al.
\newblock Experimental demonstration of memory-enhanced quantum communication.
\newblock {\em Nature}, 580(7801):60--64, 2020.

\bibitem{PhysRevLett.66.2593}
K.-J. Boller, A.~Imamo\ifmmode~\breve{g}\else \u{g}\fi{}lu, and S.~E. Harris.
\newblock Observation of electromagnetically induced transparency.
\newblock {\em Phys. Rev. Lett.}, 66:2593--2596, May 1991.

\bibitem{PhysRevLett.84.5094}
M.~Fleischhauer and M.~D. Lukin.
\newblock Dark-state polaritons in electromagnetically induced transparency.
\newblock {\em Phys. Rev. Lett.}, 84:5094--5097, May 2000.

\bibitem{PhysRevLett.120.183602}
Ya-Fen Hsiao, Pin-Ju Tsai, Hung-Shiue Chen, Sheng-Xiang Lin, Chih-Chiao Hung, Chih-Hsi Lee, Yi-Hsin Chen, Yong-Fan Chen, Ite~A. Yu, and Ying-Cheng Chen.
\newblock Highly efficient coherent optical memory based on electromagnetically induced transparency.
\newblock {\em Phys. Rev. Lett.}, 120:183602, May 2018.

\bibitem{bao2021demand}
Zenghui Bao, Zhiling Wang, Yukai Wu, Yan Li, Cheng Ma, Yipu Song, Hongyi Zhang, and Luming Duan.
\newblock On-demand storage and retrieval of microwave photons using a superconducting multiresonator quantum memory.
\newblock {\em Physical Review Letters}, 127(1):010503, 2021.

\bibitem{liu2017electromagnetically}
Yong-Chun Liu, Bei-Bei Li, and Yun-Feng Xiao.
\newblock Electromagnetically induced transparency in optical microcavities.
\newblock {\em Nanophotonics}, 6(5):789--811, 2017.

\bibitem{wang2019enhanced}
Lan Wang, Xiaoqing Guo, Yaxin Zhang, Xinlan Zhou, Lin Yuan, Ping Zhang, Shixiong Liang, Feng Lan, Hongxin Zeng, Ting Zhang, et~al.
\newblock Enhanced thz eit resonance based on the coupled electric field dropping effect within the undulated meta-surface.
\newblock {\em Nanophotonics}, 8(6):1071--1078, 2019.

\bibitem{lao2019dynamically}
Chaode Lao, Yaoyao Liang, Xianjun Wang, Haihua Fan, Faqiang Wang, Hongyun Meng, Jianping Guo, Hongzhan Liu, and Zhongchao Wei.
\newblock Dynamically tunable resonant strength in electromagnetically induced transparency (eit) analogue by hybrid metal-graphene metamaterials.
\newblock {\em Nanomaterials}, 9(2):171, 2019.

\bibitem{feng2020efficient}
L~Feng, WL~Tan, A~De, A~Menon, A~Chu, Guido Pagano, and Christopher Monroe.
\newblock Efficient ground-state cooling of large trapped-ion chains with an electromagnetically-induced-transparency tripod scheme.
\newblock {\em Physical Review Letters}, 125(5):053001, 2020.

\bibitem{jordan2019near}
Elena Jordan, Kevin~A Gilmore, Athreya Shankar, Arghavan Safavi-Naini, Justin~G Bohnet, Murray~J Holland, and John~J Bollinger.
\newblock Near ground-state cooling of two-dimensional trapped-ion crystals with more than 100 ions.
\newblock {\em Physical review letters}, 122(5):053603, 2019.

\bibitem{PhysRevLett.120.083602}
Junling Long, H.~S. Ku, Xian Wu, Xiu Gu, Russell~E. Lake, Mustafa Bal, Yu-xi Liu, and David~P. Pappas.
\newblock Electromagnetically induced transparency in circuit quantum electrodynamics with nested polariton states.
\newblock {\em Phys. Rev. Lett.}, 120:083602, Feb 2018.

\bibitem{PhysRevA.93.053838}
Qi-Chun Liu, Tie-Fu Li, Xiao-Qing Luo, Hu~Zhao, Wei Xiong, Ying-Shan Zhang, Zhen Chen, J.~S. Liu, Wei Chen, Franco Nori, J.~S. Tsai, and J.~Q. You.
\newblock Method for identifying electromagnetically induced transparency in a tunable circuit quantum electrodynamics system.
\newblock {\em Phys. Rev. A}, 93:053838, May 2016.

\bibitem{brehm2022slowing}
Jan~David Brehm, Richard Gebauer, Alexander Stehli, Alexander~N Poddubny, Oliver Sander, Hannes Rotzinger, and Alexey~V Ustinov.
\newblock Slowing down light in a qubit metamaterial.
\newblock {\em Applied Physics Letters}, 121(20), 2022.

\bibitem{wen2018reflective}
PY~Wen, AF~Kockum, H~Ian, JC~Chen, F~Nori, and I-C Hoi.
\newblock Reflective amplification without population inversion from a strongly driven superconducting qubit.
\newblock {\em Physical Review Letters}, 120(6):063603, 2018.

\bibitem{wilson2011observation}
Christopher~M Wilson, G{\"o}ran Johansson, Arsalan Pourkabirian, Michael Simoen, J~Robert Johansson, Tim Duty, Franco Nori, and Per Delsing.
\newblock Observation of the dynamical casimir effect in a superconducting circuit.
\newblock {\em nature}, 479(7373):376--379, 2011.

\bibitem{RevModPhys.89.021001}
Dibyendu Roy, C.~M. Wilson, and Ofer Firstenberg.
\newblock Colloquium: Strongly interacting photons in one-dimensional continuum.
\newblock {\em Rev. Mod. Phys.}, 89:021001, May 2017.

\bibitem{Nat.Commun.12.6383.(2021)}
Nathana{\"e}l Cottet, Haonan Xiong, Long~B. Nguyen, Yen-Hsiang Lin, and Vladimir~E. Manucharyan.
\newblock Electron shelving of a superconducting artificial atom.
\newblock {\em Nature Communications}, 12(1):6383, Nov 2021.

\bibitem{kou2018simultaneous}
A~Kou, WC~Smith, U~Vool, IM~Pop, KM~Sliwa, M~Hatridge, L~Frunzio, and MH~Devoret.
\newblock Simultaneous monitoring of fluxonium qubits in a waveguide.
\newblock {\em Physical Review Applied}, 9(6):064022, 2018.

\bibitem{brehm2021waveguide}
Jan~David Brehm, Alexander~N Poddubny, Alexander Stehli, Tim Wolz, Hannes Rotzinger, and Alexey~V Ustinov.
\newblock Waveguide bandgap engineering with an array of superconducting qubits.
\newblock {\em npj Quantum Materials}, 6(1):10, 2021.

\bibitem{wen2019large}
PY~Wen, K-T Lin, AF~Kockum, B~Suri, H~Ian, JC~Chen, SY~Mao, CC~Chiu, P~Delsing, F~Nori, et~al.
\newblock Large collective lamb shift of two distant superconducting artificial atoms.
\newblock {\em Physical Review Letters}, 123(23):233602, 2019.

\bibitem{cheng2024tuning}
Y-T Cheng, C-H Chien, K-M Hsieh, Y-H Huang, PY~Wen, W-J Lin, Y~Lu, F~Aziz, C-P Lee, K-T Lin, et~al.
\newblock Tuning atom-field interaction via phase shaping.
\newblock {\em Physical Review A}, 109(2):023705, 2024.

\bibitem{hoi2011demonstration}
Io-Chun Hoi, CM~Wilson, G{\"o}ran Johansson, Tauno Palomaki, Borja Peropadre, and Per Delsing.
\newblock Demonstration of a single-photon router in the microwave regime.
\newblock {\em Physical review letters}, 107(7):073601, 2011.

\bibitem{kannan2023demand}
Bharath Kannan, Aziza Almanakly, Youngkyu Sung, Agustin Di~Paolo, David~A Rower, Jochen Braum{\"u}ller, Alexander Melville, Bethany~M Niedzielski, Amir Karamlou, Kyle Serniak, et~al.
\newblock On-demand directional microwave photon emission using waveguide quantum electrodynamics.
\newblock {\em Nature Physics}, 19(3):394--400, 2023.

\bibitem{gheeraert2020programmable}
Nicolas Gheeraert, Shingo Kono, and Yasunobu Nakamura.
\newblock Programmable directional emitter and receiver of itinerant microwave photons in a waveguide.
\newblock {\em Physical Review A}, 102(5):053720, 2020.

\bibitem{redchenko2023tunable}
Elena~S Redchenko, Alexander~V Poshakinskiy, Riya Sett, Martin {\v{Z}}emli{\v{c}}ka, Alexander~N Poddubny, and Johannes~M Fink.
\newblock Tunable directional photon scattering from a pair of superconducting qubits.
\newblock {\em Nature Communications}, 14(1):2998, 2023.

\bibitem{zhou2023realizing}
Chao Zhou, Pinlei Lu, Matthieu Praquin, Tzu-Chiao Chien, Ryan Kaufman, Xi~Cao, Mingkang Xia, Roger~SK Mong, Wolfgang Pfaff, David Pekker, et~al.
\newblock Realizing all-to-all couplings among detachable quantum modules using a microwave quantum state router.
\newblock {\em npj Quantum Information}, 9(1):54, 2023.

\bibitem{novikov2016raman}
Sergey Novikov, T~Sweeney, JE~Robinson, SP~Premaratne, B~Suri, FC~Wellstood, and BS~Palmer.
\newblock Raman coherence in a circuit quantum electrodynamics lambda system.
\newblock {\em Nature Physics}, 12(1):75--79, 2016.

\bibitem{PhysRevResearch.5.033192}
Kai-I Chu, Wen-Te Liao, and Yung-Fu Chen.
\newblock Three-level $\mathrm{\ensuremath{\Lambda}}$-type microwave memory via parametric-modulation-induced transparency in a superconducting quantum circuit.
\newblock {\em Phys. Rev. Res.}, 5:033192, Sep 2023.

\bibitem{chiang2022tunable}
Kuan-Hsun Chiang and Yung-Fu Chen.
\newblock Tunable $\lambda$-type system made of a superconducting qubit pair.
\newblock {\em Physical Review A}, 106(2):023707, 2022.

\bibitem{PhysRevA.103.023710}
A.~M. Vadiraj, Andreas Ask, T.~G. McConkey, I.~Nsanzineza, C.~W.~Sandbo Chang, Anton~Frisk Kockum, and C.~M. Wilson.
\newblock Engineering the level structure of a giant artificial atom in waveguide quantum electrodynamics.
\newblock {\em Phys. Rev. A}, 103:023710, Feb 2021.

\bibitem{abdumalikov2010electromagnetically}
AA~Abdumalikov~Jr, O~Astafiev, Alexandre~M Zagoskin, Yu~A Pashkin, Y~Nakamura, and Jaw~Shen Tsai.
\newblock Electromagnetically induced transparency on a single artificial atom.
\newblock {\em Physical review letters}, 104(19):193601, 2010.

\bibitem{vadiraj2021engineering}
AM~Vadiraj, Andreas Ask, TG~McConkey, I~Nsanzineza, CW~Sandbo Chang, Anton~Frisk Kockum, and CM~Wilson.
\newblock Engineering the level structure of a giant artificial atom in waveguide quantum electrodynamics.
\newblock {\em Physical Review A}, 103(2):023710, 2021.

\bibitem{fluxoniumQ}
Long~B. Nguyen, Yen-Hsiang Lin, Aaron Somoroff, Raymond Mencia, Nicholas Grabon, and Vladimir~E. Manucharyan.
\newblock High-coherence fluxonium qubit.
\newblock {\em Phys. Rev. X}, 9:041041, Nov 2019.

\bibitem{DolanBridge}
G.~J. Dolan.
\newblock {Offset masks for lift‐off photoprocessing}.
\newblock {\em Applied Physics Letters}, 31(5):337--339, 08 2008.

\bibitem{superinductance}
Vladimir~E. Manucharyan, Nicholas~A. Masluk, Archana Kamal, Jens Koch, Leonid~I. Glazman, and Michel~H. Devoret.
\newblock Evidence for coherent quantum phase slips across a josephson junction array.
\newblock {\em Phys. Rev. B}, 85:024521, Jan 2012.

\bibitem{PhysRevLett.103.217004}
Jens Koch, V.~Manucharyan, M.~H. Devoret, and L.~I. Glazman.
\newblock Charging effects in the inductively shunted josephson junction.
\newblock {\em Phys. Rev. Lett.}, 103:217004, Nov 2009.

\bibitem{SuppMaterial}
Supplementary material.

\bibitem{Lindblad1976}
G.~Lindblad.
\newblock On the generators of quantum dynamical semigroups.
\newblock {\em Communications in Mathematical Physics}, 48(2):119--130, Jun 1976.

\bibitem{PhysRevLett.130.267001}
Aaron Somoroff, Quentin Ficheux, Raymond~A. Mencia, Haonan Xiong, Roman Kuzmin, and Vladimir~E. Manucharyan.
\newblock Millisecond coherence in a superconducting qubit.
\newblock {\em Phys. Rev. Lett.}, 130:267001, Jun 2023.

\bibitem{wiegand2021ultimate}
Emely Wiegand, Ping-Yi Wen, Per Delsing, Io-Chun Hoi, and Anton~Frisk Kockum.
\newblock Ultimate quantum limit for amplification: a single atom in front of a mirror.
\newblock {\em New Journal of Physics}, 23(4):043048, 2021.

\bibitem{lu2021propagating}
Yong Lu, Ingrid Strandberg, Fernando Quijandr{\'\i}a, G{\"o}ran Johansson, Simone Gasparinetti, and Per Delsing.
\newblock Propagating wigner-negative states generated from the steady-state emission of a superconducting qubit.
\newblock {\em Physical Review Letters}, 126(25):253602, 2021.

\bibitem{johansson2012qutip}
J~Robert Johansson, Paul~D Nation, and Franco Nori.
\newblock Qutip: An open-source python framework for the dynamics of open quantum systems.
\newblock {\em Computer physics communications}, 183(8):1760--1772, 2012.

\bibitem{PhysRevA.88.043806}
Kevin Lalumi\`ere, Barry~C. Sanders, A.~F. van Loo, A.~Fedorov, A.~Wallraff, and A.~Blais.
\newblock Input-output theory for waveguide qed with an ensemble of inhomogeneous atoms.
\newblock {\em Phys. Rev. A}, 88:043806, Oct 2013.

\bibitem{PhysRevLett.124.240402}
Gustav Andersson, Maria~K. Ekstr\"om, and Per Delsing.
\newblock Electromagnetically induced acoustic transparency with a superconducting circuit.
\newblock {\em Phys. Rev. Lett.}, 124:240402, Jun 2020.

\bibitem{cheng2025group}
Y-T Cheng, K-M Hsieh, B-Y Wu, ZQ~Niu, F~Aziz, Y-H Huang, PY~Wen, K-T Lin, Y-H Lin, JC~Chen, et~al.
\newblock Group delay controlled by the decoherence of a single artificial atom.
\newblock {\em Physical Review Letters}, 135(7):073601, 2025.

\bibitem{chu2025slow}
Kai-I Chu, Xiao-Cheng Lu, Kuan-Hsun Chiang, Yen-Hsiang Lin, Chii-Dong Chen, Ite~A Yu, Wen-Te Liao, and Yung-Fu Chen.
\newblock Slow and stored light via electromagnetically induced transparency using a $\lambda$-type superconducting artificial atom.
\newblock {\em Physical Review Research}, 7(1):L012015, 2025.

\bibitem{mencia2024integer}
Raymond~A Mencia, Wei-Ju Lin, Hyunheung Cho, Maxim~G Vavilov, and Vladimir~E Manucharyan.
\newblock Integer fluxonium qubit.
\newblock {\em PRX Quantum}, 5(4):040318, 2024.

\bibitem{ardati2024using}
Wa{\"e}l Ardati, S{\'e}bastien L{\'e}ger, Shelender Kumar, Vishnu~Narayanan Suresh, Dorian Nicolas, Cyril Mori, Francesca D’Esposito, Tereza Vakhtel, Olivier Buisson, Quentin Ficheux, et~al.
\newblock Using bifluxon tunneling to protect the fluxonium qubit.
\newblock {\em Physical Review X}, 14(4):041014, 2024.

\end{thebibliography}


\begin{thebibliography}{4}%
\makeatletter
\providecommand \@ifxundefined [1]{%
 \@ifx{#1\undefined}
}%
\providecommand \@ifnum [1]{%
 \ifnum #1\expandafter \@firstoftwo
 \else \expandafter \@secondoftwo
 \fi
}%
\providecommand \@ifx [1]{%
 \ifx #1\expandafter \@firstoftwo
 \else \expandafter \@secondoftwo
 \fi
}%
\providecommand \natexlab [1]{#1}%
\providecommand \enquote  [1]{``#1''}%
\providecommand \bibnamefont  [1]{#1}%
\providecommand \bibfnamefont [1]{#1}%
\providecommand \citenamefont [1]{#1}%
\providecommand \href@noop [0]{\@secondoftwo}%
\providecommand \href [0]{\begingroup \@sanitize@url \@href}%
\providecommand \@href[1]{\@@startlink{#1}\@@href}%
\providecommand \@@href[1]{\endgroup#1\@@endlink}%
\providecommand \@sanitize@url [0]{\catcode `\\12\catcode `\$12\catcode `\&12\catcode `\#12\catcode `\^12\catcode `\_12\catcode `\%12\relax}%
\providecommand \@@startlink[1]{}%
\providecommand \@@endlink[0]{}%
\providecommand \url  [0]{\begingroup\@sanitize@url \@url }%
\providecommand \@url [1]{\endgroup\@href {#1}{\urlprefix }}%
\providecommand \urlprefix  [0]{URL }%
\providecommand \Eprint [0]{\href }%
\providecommand \doibase [0]{http://dx.doi.org/}%
\providecommand \selectlanguage [0]{\@gobble}%
\providecommand \bibinfo  [0]{\@secondoftwo}%
\providecommand \bibfield  [0]{\@secondoftwo}%
\providecommand \translation [1]{[#1]}%
\providecommand \BibitemOpen [0]{}%
\providecommand \bibitemStop [0]{}%
\providecommand \bibitemNoStop [0]{.\EOS\space}%
\providecommand \EOS [0]{\spacefactor3000\relax}%
\providecommand \BibitemShut  [1]{\csname bibitem#1\endcsname}%
\let\auto@bib@innerbib\@empty
\bibitem [{\citenamefont {Wen}\ \emph {et~al.}(2018)\citenamefont {Wen}, \citenamefont {Kockum}, \citenamefont {Ian}, \citenamefont {Chen}, \citenamefont {Nori},\ and\ \citenamefont {Hoi}}]{PhysRevLett.120.063603}%
  \BibitemOpen
  \bibfield  {author} {\bibinfo {author} {\bibfnamefont {P.~Y.}\ \bibnamefont {Wen}}, \bibinfo {author} {\bibfnamefont {A.~F.}\ \bibnamefont {Kockum}}, \bibinfo {author} {\bibfnamefont {H.}~\bibnamefont {Ian}}, \bibinfo {author} {\bibfnamefont {J.~C.}\ \bibnamefont {Chen}}, \bibinfo {author} {\bibfnamefont {F.}~\bibnamefont {Nori}}, \ and\ \bibinfo {author} {\bibfnamefont {I.-C.}\ \bibnamefont {Hoi}},\ }\bibfield  {title} {\enquote {\bibinfo {title} {Reflective amplification without population inversion from a strongly driven superconducting qubit},}\ }\href {\doibase 10.1103/PhysRevLett.120.063603} {\bibfield  {journal} {\bibinfo  {journal} {Phys. Rev. Lett.}\ }\textbf {\bibinfo {volume} {120}},\ \bibinfo {pages} {063603} (\bibinfo {year} {2018})}\BibitemShut {NoStop}%
\bibitem [{\citenamefont {Peng}\ \emph {et~al.}(2014)\citenamefont {Peng}, \citenamefont {{\"O}zdemir}, \citenamefont {Chen}, \citenamefont {Nori},\ and\ \citenamefont {Yang}}]{Nat.Commu.5.5082.(2014)}%
  \BibitemOpen
  \bibfield  {author} {\bibinfo {author} {\bibfnamefont {B.}~\bibnamefont {Peng}}, \bibinfo {author} {\bibfnamefont {{\c{S}}.~K.}\ \bibnamefont {{\"O}zdemir}}, \bibinfo {author} {\bibfnamefont {W.}~\bibnamefont {Chen}}, \bibinfo {author} {\bibfnamefont {F.}~\bibnamefont {Nori}}, \ and\ \bibinfo {author} {\bibfnamefont {L.}~\bibnamefont {Yang}},\ }\bibfield  {title} {\enquote {\bibinfo {title} {What is and what is not electromagnetically induced transparency in whispering-gallery microcavities},}\ }\href@noop {} {\bibfield  {journal} {\bibinfo  {journal} {Nature Communications}\ }\textbf {\bibinfo {volume} {5}},\ \bibinfo {pages} {5082} (\bibinfo {year} {2014})}\BibitemShut {NoStop}%
\bibitem [{\citenamefont {Anisimov}, \citenamefont {Dowling},\ and\ \citenamefont {Sanders}(2011)}]{PhysRevLett.107.163604}%
  \BibitemOpen
  \bibfield  {author} {\bibinfo {author} {\bibfnamefont {P.~M.}\ \bibnamefont {Anisimov}}, \bibinfo {author} {\bibfnamefont {J.~P.}\ \bibnamefont {Dowling}}, \ and\ \bibinfo {author} {\bibfnamefont {B.~C.}\ \bibnamefont {Sanders}},\ }\bibfield  {title} {\enquote {\bibinfo {title} {Objectively discerning autler-townes splitting from electromagnetically induced transparency},}\ }\href {\doibase 10.1103/PhysRevLett.107.163604} {\bibfield  {journal} {\bibinfo  {journal} {Phys. Rev. Lett.}\ }\textbf {\bibinfo {volume} {107}},\ \bibinfo {pages} {163604} (\bibinfo {year} {2011})}\BibitemShut {NoStop}%
\bibitem [{\citenamefont {Sun}\ \emph {et~al.}(2014)\citenamefont {Sun}, \citenamefont {Liu}, \citenamefont {Ian}, \citenamefont {You}, \citenamefont {Il'ichev},\ and\ \citenamefont {Nori}}]{PhysRevA.89.063822}%
  \BibitemOpen
  \bibfield  {author} {\bibinfo {author} {\bibfnamefont {H.-C.}\ \bibnamefont {Sun}}, \bibinfo {author} {\bibfnamefont {Y.-x.}\ \bibnamefont {Liu}}, \bibinfo {author} {\bibfnamefont {H.}~\bibnamefont {Ian}}, \bibinfo {author} {\bibfnamefont {J.~Q.}\ \bibnamefont {You}}, \bibinfo {author} {\bibfnamefont {E.}~\bibnamefont {Il'ichev}}, \ and\ \bibinfo {author} {\bibfnamefont {F.}~\bibnamefont {Nori}},\ }\bibfield  {title} {\enquote {\bibinfo {title} {Electromagnetically induced transparency and autler-townes splitting in superconducting flux quantum circuits},}\ }\href {\doibase 10.1103/PhysRevA.89.063822} {\bibfield  {journal} {\bibinfo  {journal} {Phys. Rev. A}\ }\textbf {\bibinfo {volume} {89}},\ \bibinfo {pages} {063822} (\bibinfo {year} {2014})}\BibitemShut {NoStop}%
\end{thebibliography}%

\end{document}


\title[]{Supplementary Material for ``Slowing and Storing Light in a Single Superconducting Fluxonium Artificial Atom''}
\author{Ching-Yeh Chen}
\affiliation{Department of Physics, National Tsing Hua University, Hsinchu 30013, Taiwan}

\author{Shih-Wei Lin}
\affiliation{Department of Physics, National Tsing Hua University, Hsinchu 30013, Taiwan}

\author{Ching-Ping Lee}
\affiliation{Department of Physics, National Tsing Hua University, Hsinchu 30013, Taiwan}

\author{J.~C.~Chen}
\affiliation{Department of Physics, National Tsing Hua University, Hsinchu 30013, Taiwan}

\author{I.-C.~Hoi}
\affiliation{Department of Physics, National Tsing Hua University, Hsinchu 30013, Taiwan}
\affiliation{Department of Physics, City University of Hong Kong, Kowloon, Hong Kong SAR 999077, China}

\author{Yen-Hsiang Lin}
\affiliation{Department of Physics, National Tsing Hua University, Hsinchu 30013, Taiwan}
\affiliation{Taiwan Semiconductor Research Institute, Hsinchu 300091, Taiwan}
\maketitle


\maketitle

\tableofcontents

\renewcommand{\thefigure}{S\arabic{figure}}
\renewcommand{\thesection}{S\arabic{section}}
\renewcommand{\theequation}{S\arabic{equation}}
\renewcommand{\thetable}{S\arabic{table}}
\renewcommand{\bibnumfmt}[1]{[S#1]}
\renewcommand{\citenumfont}[1]{S#1}


\newpage

\section{setup}
\begin{figure}
\includegraphics[scale=0.2]{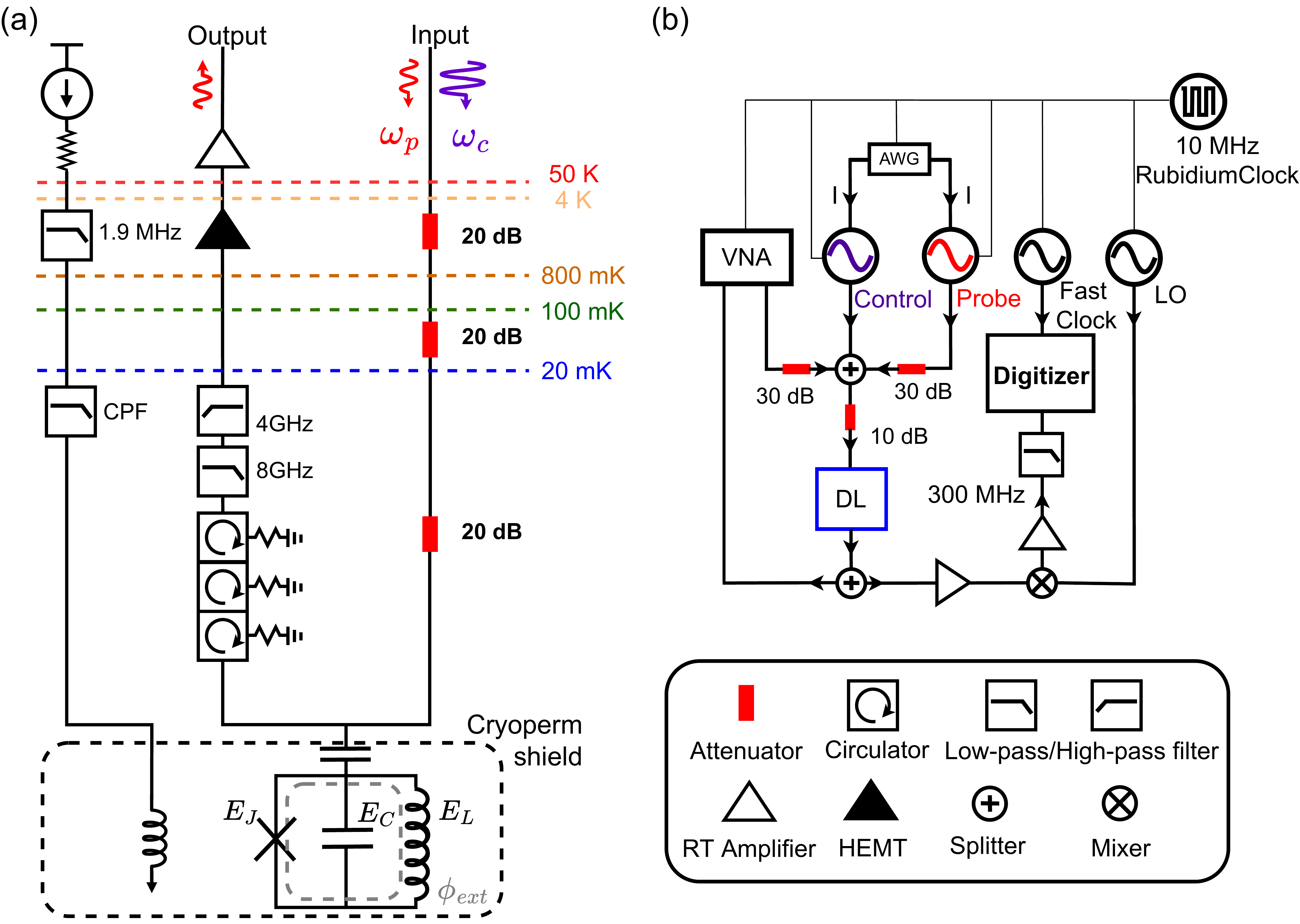}
\caption{\label{fig:pdfart}  The schematics of the full experimental measurement setup. (a) The device is placed at the bottom of the dilution refrigerator operating at the base temperature 20 mK (blue dashed line). We control the qubit frequency by sending in the directional current (DC) source. (b) The room-temperature measurement setup includes a vector network analyzer (VNA) for spectroscopy, which is connected in parallel with the time-domain system composed of a digitizer, an arbitrary waveform generator (AWG), and radio-frequency (rf) sources. The signal from the output port of the refrigerator is mixed with a local oscillator (LO) for down-conversion before being recorded by the digitizer. }
\label{setup}
\end{figure}

In Fig. \ref{setup}(a), we cool down the device in the dilution refrigerator with a base temperature of approximately 20 mK in the mixing chamber. To reduce Johnson thermal noise from the room temperature environment, attenuators (depicted as red rectangles) are strategically thermalized along the input coaxial lines at each fridge stage. The output signal is then routed through triple isolators positioned at the mixing chamber stage. Following this, it passes through an amplifier chain comprising a HEMT amplifier at the 4K fridge stage and a low-noise amplifier at room temperature. The 20 mK space is safeguarded by a radiation shield, and the device resides within a single layer of a cylindrical Cryoperm magnetic shield. Within this shield, a superconducting wire is employed to manipulate the qubit frequency by applying a magnetic field through a DC current source, while a copper waveguide integrates with the fluxonium qubit. The spectroscopic measurement operates concurrently with the time-domain measurement system. The spectrum is acquired using a Vector Network Analyzer (VNA), with the radio frequency (RF) generator providing the control field. In the time-domain measurement, an arbitrary waveform generator (AWG) and a Digitizer are utilized. The pulse is generated by the AWG, up-converted with RF generators (Probe and Control) signals, and subsequently, the signal from the Dilution Fridge is down-converted by the local oscillator(LO) before being measured by the Digitizer.

\section{qubit parameters}
\begin{figure}
\includegraphics[scale=0.7]{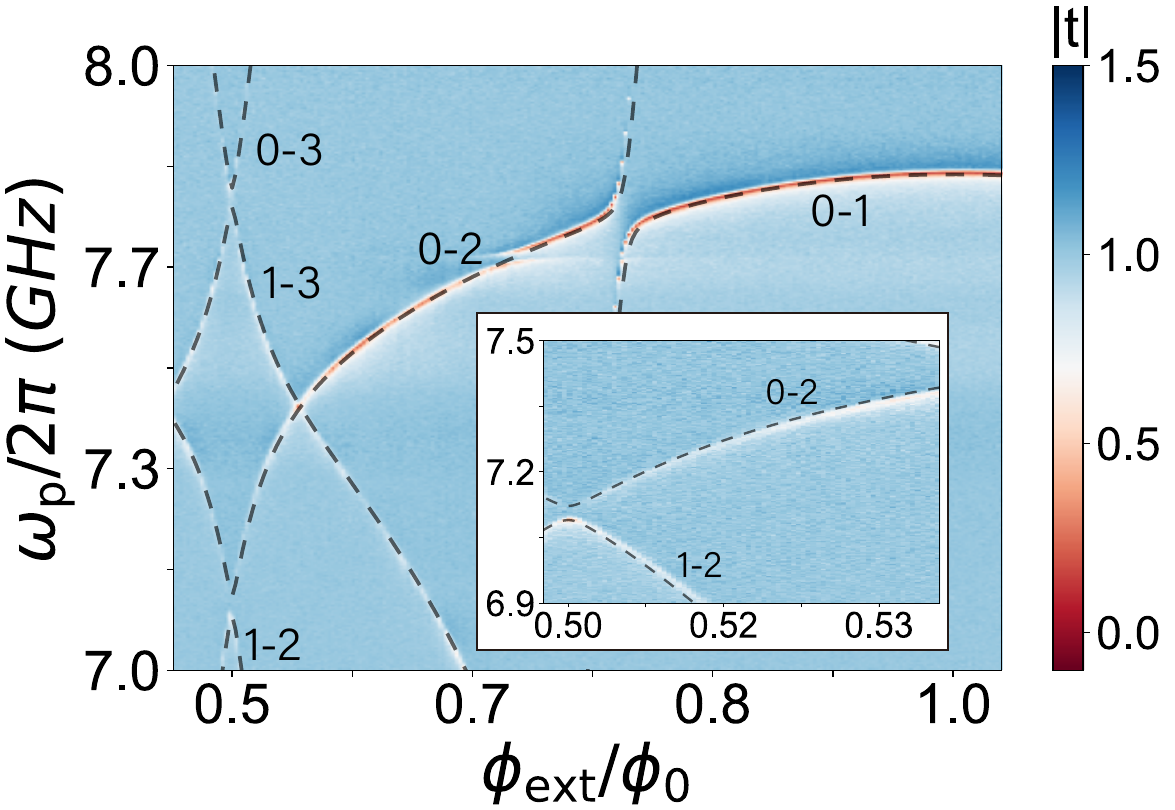}
\caption{\label{fig:pdfart} One-tone spectroscopy of the fluxonium qubit. The transmission coefficient, $|t|$, as a function of the probing frequency, $\omega_{\rm{p}}$, and $\phi_{\rm{ext}}$. Dashed lines indicate the fitting result to the transition frequency by the Hamiltonian (1). The inset shows the zooming in result closing to the sweep spot, where we bias at $\phi_{\rm{ext}}/\phi_0=0.53$.}
\label{flux_dp}
\end{figure}

The data were obtained through one-tone spectroscopy\cite{PhysRevLett.120.063603}. The fitting line in Fig. \ref{flux_dp} illustrates the numerical diagonalization outcome of Hamiltonian (1), with $E_{J}=9.041$ GHz, $E_{C}=0.995$ GHz, $E_{L}=0.807$ GHz, and flux-to-coil current conversion as adjustable parameters.



\section{Three level Lambda system in the Rotating Wave}

To elucidate the dynamics of our system, we focus on the three lowest states, denoted as $|0\rangle$, $|1\rangle$, and $|2\rangle$, with corresponding projectors $\sigma_{ij}=|i\rangle\langle j|$. In constructing a comprehensive three-level system, we assign $\omega_0$, $\omega_1$ and $\omega_2$ as the energies of $|0\rangle$, $|1\rangle$ and $|2\rangle$, respectively.

Our approach involves applying the control field $\omega_{\rm{c}}$ to the $|1\rangle$-$|2\rangle$ transition and the weak probe $\omega_{\rm{p}}$ to the $|0\rangle$-$|2\rangle$ transition. This leads to the following Hamiltonian:

\begin{equation}
\begin{split}
{H} = &\omega_{0}\sigma_{00}+\omega_{1}\sigma_{11}+\omega_{2}\sigma_{22}-\\
&\Omega_{\rm{p}}\cos{\omega_{\rm{p}}t}(\sigma_{02}+\sigma_{20})-\Omega_{\rm{c}}\cos{\omega_{\rm{c}}t}(\sigma_{12}+\sigma_{21})
\end{split}
\end{equation}

The time dependence can be eliminated by transitioning into a rotating frame through the application of a unitary transformation. This results in the rotated Hamiltonian.
\begin{equation}
\begin{split}
{U} = \exp\left[{it(\omega_{0}\sigma_{00}+(\omega_{0}+\omega_{\rm{p}}-\omega_{\rm{c}})+\sigma_{11}(\omega_{\rm{p}}-\omega_{\rm{c}})+\sigma_{22}(\omega_{0}+\omega_{\rm{p}})}\right]
\end{split}
\end{equation}

\begin{equation}
\begin{split}
\widetilde{H} = UH_aU + i\partial_tUU^\dag
\end{split}
\end{equation}

By choosing the zero energy state as $|0\rangle$, applying the Rotating Wave Approximation (RWA), and introducing $\Delta_{\rm{c}}=\omega_{21}-\omega_{\rm{c}}$ and $\Delta_{\rm{p}}=\omega_{20}-\omega_{\rm{p}}$, we derive a time-independent Hamiltonian. This is accomplished by incorporating two driving strengths, denoted as $\Omega_{\rm{c}}$ and $\Omega_{\rm{p}}$.

\begin{equation}
\begin{split}
\widetilde{H} = &-\frac{\Omega_{\rm{p}}}{2}(\sigma_{02}+\sigma_{20})-\frac{\Omega_{\rm{c}}}{2}(\sigma_{12}+\sigma_{21})+(\Delta_{\rm{p}}-\Delta_{\rm{c}})\sigma_{11}+\Delta_{\rm{p}}\sigma_{22}
\end{split}
\end{equation}




\section{Akaike Information Criterion}

\begin{figure}
    \includegraphics[scale=0.43]{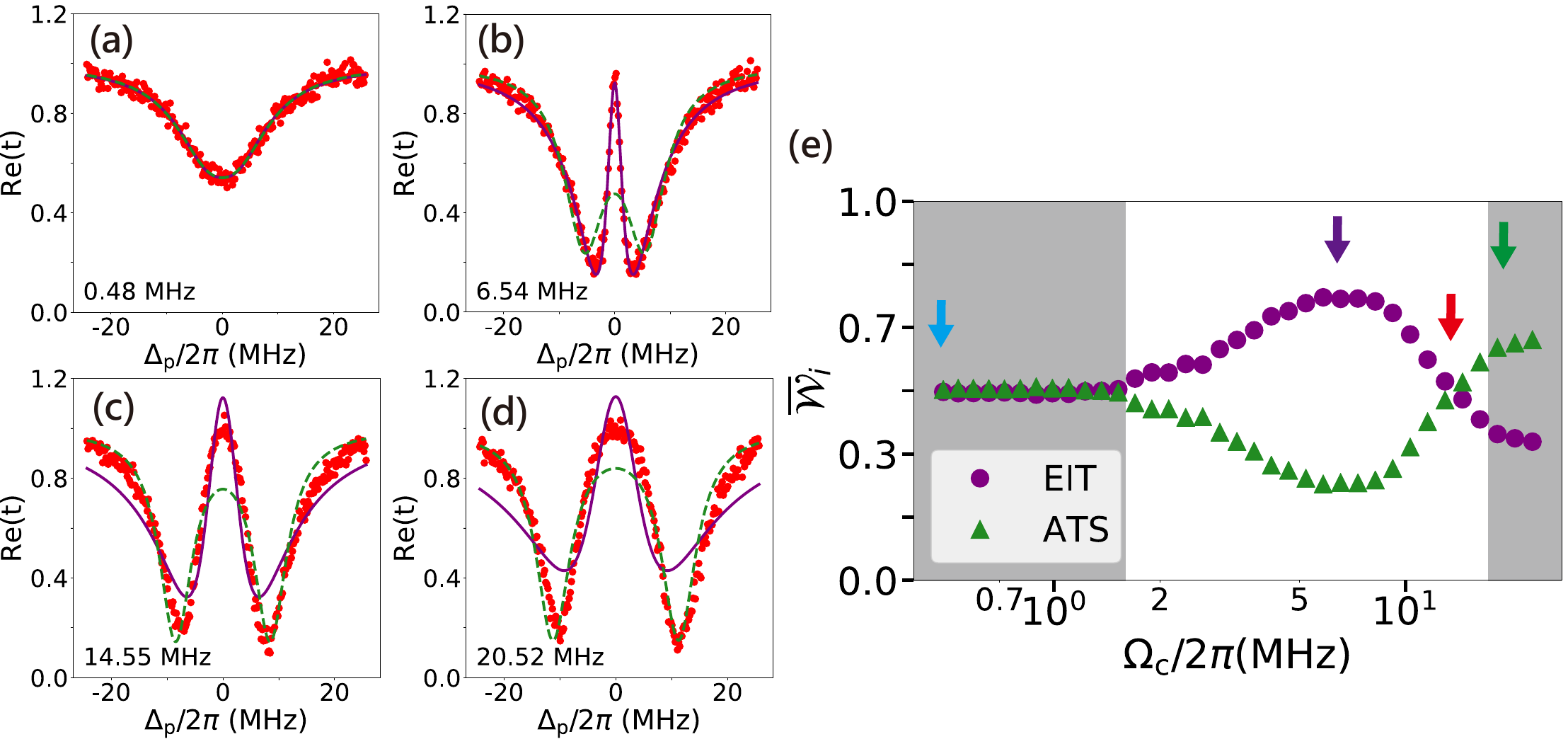}
    \caption{\label{fig:pdfart} 
Akaike Information Criterion (AIC). (a)-(d) Comparison of the measured real part of the transmission coefficient (red points) under four different $\Omega_{\rm{c}}/2\pi: 0.48, 6.54, 14.55$, and $ 20.52\,\, \rm{MHz}$, fitted with the ATS and EIT models, depicted as green dashed lines and purple lines, respectively. (e) The AIC per-point weight for both ATS and EIT models obtained from fitting the experimental data. The corresponding $\Omega_{\rm{c}}$ values of (a)-(d) are indicated by arrows from low to high value of the $\Omega_{\rm{c}}$. Blue and red arrows delineate the transition domain, with the purple (green) arrow indicating the dominance of the maximum EIT (ATS) model.}
    \label{AIC}
    \end{figure}

In our extended analysis, we employ the Akaike Information Criterion (AIC)\cite{Nat.Commu.5.5082.(2014)}proposed to discriminate and distinguish the Electromagnetically Induced Transparency (EIT) regime from the Autler-Townes Splitting (ATS) regime within the driven three-level system. The AIC offers a method to select the optimal model from a set of models, relying on the Kullback–Leibler (K–L) distance, which quantifies the divergence between the model and the truth. The superior model exhibits minimal information loss and, consequently, the smallest K-L distance. The model $\lambda_{i}$ fitting data points$x=x_{1},x_{2},...,x_{n}$ with n number of data points and $k_{i}$ parameters, and analysis as $I_{i}=-2logL_{i}+2k_{i}$, where $L_{i}$ represents the maximum likelihood for the candidate model, accounting for the penalty associated with the number of parameters used in the fitting.

\begin{equation}
\begin{split}
{T_{\rm{EIT}}}={1-\frac{C_{1}}{\omega^{2}+\Gamma_{1}^{2}}+\frac{C_{2}}{\omega^{2}+\Gamma_{2}^{2}}}
\label{T_EIT}
\end{split}
\end{equation}

\begin{equation}
\begin{split}
{T_{\rm{ATS}}}={1-\frac{C}{(\omega-\delta)^{2}+\Gamma^{2}}-\frac{C}{(\omega+\delta)^{2}+\Gamma^{2}}}
\label{T_ATS}
\end{split}
\end{equation}

We individually fit the EIT and ATS regimes with transmission $\rm{Re(t)}$ and rewrite into $T_{\rm{EIT}}$ and $T_{\rm{ATS}}$. In Fig. \ref{AIC}(a)-(d), we directly compare our experimental result with the EIT model in Eq.\ref{T_EIT} and the ATS model in Eq.\ref{T_ATS}. For a weak control field, the observed transmission spectrum starts to split, fitting very well with the Electromagnetically Induced Transparency (EIT) model and deviates noticeably from the Autler-Townes Splitting (ATS) model, shown in Fig. \ref{AIC}(b). However, when the control field has moderate strength, the transmission spectrum deviates significantly from both the EIT and ATS models [Fig. \ref{AIC}(c) and Fig. \ref{AIC}(d)]. In Fig. \ref{AIC}(a), both models fit well, indicating that neither model dominates.
 Because of least-squares and the presence of technical noise in the experiments, we show the per-point (mean) AIC weight in Eq.\ref{weight} and Eq.\ref{w_ats}, where $\bar I_{i}=I_{i}/n$
\begin{equation}
\begin{split}
{\bar w_{\rm{EIT}}}={\frac{\rm{exp}(-\frac{1}{2}\bar I_{\rm{EIT}})}{\rm{exp}(-\frac{1}{2}\bar I_{\rm{EIT}})+\rm{exp}(-\frac{1}{2}\bar I_{\rm{ATS}})}}
\label{weight}
\end{split}
\end{equation}

\begin{equation}
\begin{split}
{\bar w_{\rm{ATS}}}={1-\bar w_{\rm{EIT}}}
\label{w_ats}
\end{split}
\end{equation}

 In Fig.\ref{AIC}(e)\cite{PhysRevLett.107.163604}. The crossing point at the AIC, which points out the $\Omega_{\rm{AIC}}$, corresponds to the $\Omega_{\rm{c}}=14.55\,\,\rm{MHz}$. Comparing the $\Omega_{\rm{EIT}}$ and $\Omega_{\rm{AIC}}$, $\Omega_{\rm{EIT}}$ is close to the maximum per-point weight, and the $\Omega_{AIC}$ shows the crossing point of the EIT and the ATS\cite{PhysRevA.89.063822}.

\bibliography{sup}